# Measurement of swelling-induced residual stress in ion implanted SiC, and its effect on micromechanical properties

Alexander J. Leide[a1*], Richard I. Todd[a], and David E. J. Armstrong[a]

[a] *Department of Materials, University of Oxford*
Parks Road, Oxford, OX1 3PH, UK

email: *alex.leide@bristol.ac.uk*

*Abstract.* Ion implantation is widely used as a surrogate for neutron irradiation in the investigation of radiation damage on the properties of materials. Due to the small depth of damage, micromechanical methods must be used to extract material properties. In this work, nanoindentation has been applied to ion irradiated silicon carbide to extract radiation-induced hardening. Residual stress is evaluated using HR-EBSD, AFM swelling measurements, and a novel microcantilever relaxation technique coupled with finite element modelling. Large compressive residual stresses of several GPa are found in the irradiated material, which contribute to the significant hardening observed in nanoindentation measurements. The origin of these residual stresses and the associated hardening is the unirradiated substrate which constrains radiation swelling. Comparisons with other materials susceptible to irradiation swelling show that this effect should not be neglected in studying the effects of ion irradiation damage on mechanical properties. This constraint may also be influencing fundamental radiation defects. This has significant implications for the suitability of ion implantation as a surrogate for neutron irradiations. These results demonstrate the significance of swelling-induced residual stresses in nuclear reactor components, and the impact on structural integrity of reactor components.

---

[1] Present address: School of Physics, University of Bristol, Tyndall Avenue, BS8 1TL, UK

**Introduction**

Silicon carbide (SiC) is a structural ceramic material useful in extreme environments, primarily aerospace and nuclear applications, because of its excellent high-temperature properties, including creep resistance, high strength at elevated temperatures, corrosion resistance and general chemical inertness, high thermal conductivity, and low thermal expansion coefficient [1–3]. SiC is desirable for applications as fission fuel cladding, or as a component of the blanket, first wall, or divertor of fusion reactors due to its low neutron absorption cross-section, low level of long-lived radioisotopes, and stability under high temperature-high dose neutron irradiation [1,4–9].

For nuclear applications, a thorough understanding of radiation defects and their effects on material properties is required to evaluate the suitability of a material for its application, and to predict the evolution of its properties over time. To accelerate radiation damage processes and material investigations, ion implantation is commonly used as a surrogate for neutron irradiation [10,11]. It allows displacement damage to be introduced to a material in controlled conditions in a matter of hours as compared to many days for comparable damage in a nuclear fission reactor. Additionally, it does not introduce radiological hazards due to sample activation, avoiding the requirements for specialist "active" laboratories, sample cooling, and remote handling.

Due to the shallow damage layer introduced by ion implantation (a few microns), evaluating the material properties requires miniaturisation of testing techniques. An array of micromechanical techniques have been developed which can be applied to materials on the scale of an ion implanted layer: micropillar compression [12,13], microcantilever bending [14,15], and nanoindentation [16]. Due to the non-uniform damage profile in ion implantations, it is difficult to extract mechanical properties as a function of damage as the indentation plastic zone interacts with a range of radiation doses, although this can be accounted for using the method of Kareer *et al.*[17]. Damage gradients can somewhat be accounted for in experimental design. Proton irradiation has a high penetration depth and a flatter damage layer before the Bragg peak, but damage in this region is limited to low dpa unless long irradiations are carried out [10]. Multi-energy ion implantations can create a flattened damage profile, but implanted ions will remain in this region [14].

As ceramics typically fail from a flaw, statistical macroscopic testing is important to understand whether a component and manufacturing process is suitable for its intended application. For nuclear applications this necessitates neutron irradiation of many specimens for statistical strengths. Where a lab has access to neutron irradiated silicon carbide, they tend to focus on macroscopic testing. From a material design and improvement perspective, micromechanical testing is vital to understand microstructural features and their influence on macroscopic properties, and as an input for multi-scale modelling. This appears to have been neglected in favour of macroscopic tests, and there is little nanoindentation of neutron irradiated SiC in the literature to compare ion irradiations to, especially of single crystals of 6H-SiC which are of interest for fundamental studies, removing the influence of microstructure.

Three researchers have published nanoindentation results from neutron irradiated monolithic SiC. Osborne *et al.* [18] and Nogami *et al.* [19,20] used CVD 3C-SiC samples as a surrogate for the matrix of CVI $SiC_f$/SiC composites while Chen *et al.* [21] used 6H-SiC single crystals as part of their comparison with ion implantation. Osborne *et al.* found that hardening saturated at ~8% by ~1 dpa but was higher at 500 °C due to growth of interstitial dislocation loops. Nogami *et al.* meanwhile found no dependence on hardening with temperature above 100 °C, with hardening of ~10% in their experiments. Chen *et al.* observed no hardening in their neutron irradiated samples at 0.1 and 0.2 dpa at ~50 °C. It is worth pointing out that Osborne's and Nogami's works are from the early days of nanoindentation. Nogami's measurements come with very large standard deviation error bars which significantly overlap the unirradiated hardness values.

Chen *et al.*'s recent results are particularly valuable as they compare neutron and ion irradiation at similar nominal temperatures and doses in 6H-SiC single crystal [21]. Carbon and silicon ion irradiations at the same temperature and dose as the neutron irradiation cause significant hardening: +14.3% and +12.5% for silicon and carbon at 0.1 dpa, and +11.8% and +9% at 0.2 dpa. As these experiments are by the same group, the nanoindentation and analysis methodology should therefore be consistent, whereas the procedures for nanoindentation can vary between researchers making comparisons unreliable.

Most ion irradiations of 6H-SiC have been conducted at room temperature, with these showing a hardening between 13% - 20.5% [22–24]. Of the elevated temperature ion irradiations, Su *et al.* [25] show 16.7 % hardening after 3.16 dpa, and 14.7% after 0.95 dpa at 600 °C using argon ions. Kerbiriou *et al.* implanted gold ions into 6H-SiC single crystals at room temperature and 400 °C and found hardening saturated at 9.8% after ~0.15 dpa and stayed approximately at that hardness without amorphising up to their highest dose of 27 dpa at 400 °C [26]. At room temperature, hardening after ion implantation below the amorphisation dose is ~12.5%. Their unirradiated nanoindentation hardness value of 48 GPa is significantly higher than 38 GPa found by most other researchers. Kerbiriou *et al.* also measured out of plane swelling using atomic force microscopy, finding expansions between 2.2% and 3.8% increasing with dose [26].

Existing research appears to show that ion irradiations cause more hardening than neutron irradiations (especially at low doses). A more thorough understanding of the mechanism behind large hardening in ion implanted layers of silicon carbide is required. This is both for evaluating the use of ion implantation for simulating neutron irradiation, and for understanding the fundamental radiation response of silicon carbide. The use of single crystals removes the effect of microstructure on radiation damage and mechanical properties, giving a better understanding of the intrinsic material response to radiation damage.

**Methods**

A pre-polished sample of 6H-SiC single crystal, with surfaces parallel to the (0001) basal plane was purchased from Pi-Kem Ltd (Tamworth, UK). Raman spectra of these single crystals demonstrate no pre-existing stress in these specimens. Ion implantation was carried out at the Surrey Ion Beam Centre, UK using the 2 MV Van de Graaf accelerator. Samples were clamped to a heated stage using washers to blank part of the specimen from the ion beam. The stage was held at 300 °C (±5 °C) in a vacuum of ~1×10$^{-6}$ mbar. The sample was implanted with neon ions at three energies (1450 keV, 720 keV, and 350 keV) to create a flattened damage profile within the approximate plastic zone of nanoindentations (Figure 1). Neon ions were chosen to avoid any chemical effects from the implanted ions, while producing a similar damage profile to our other work using silicon ions [27]. Self-ions are typically chosen for implantation of metals to avoid chemical effects; however literature and our own work suggests implanting SiC with silicon or carbon ions may influence defect types compared to neutron irradiations [28]. Chemical defects are important in SiC, so influencing defect chemistry with ion implantation should be avoided [29]. Displacements per atom (dpa) was calculated using the Stopping and Range of Ions in Matter (SRIM) Monte Carlo code with the quick Kinchin-Pease model [30,31]. Displacement energies for silicon and carbon were 35 eV and 21 eV respectively, with binding energies set to 0 eV [32]. Target density was set to 3.21 g/cm$^3$. The peak nominal damage is ~2.5 dpa with a mean of 1.8 dpa.

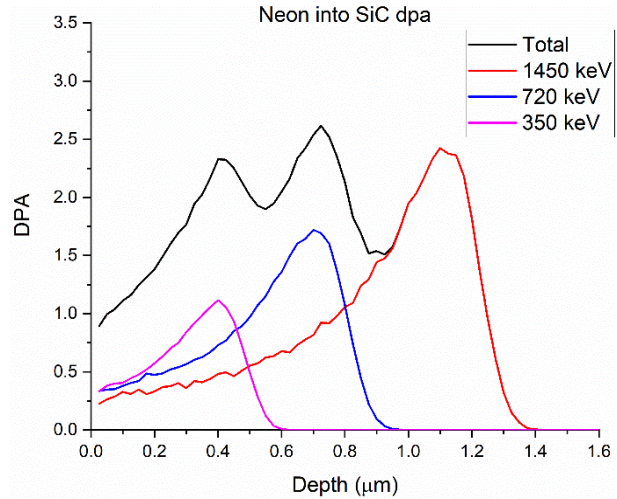

*Figure 1: Damage profile of neon ion implantation into 6H-SiC, as calculated from SRIM* [30]

Nanoindentation was carried out using an MTS Nanoindenter XP with a diamond Berkovich tip. The continuous stiffness method (CSM) was used to calculate mechanical properties. The CSM harmonic displacement was 2 nm with a frequency of 45 Hz and a strain rate of 0.05 s$^{-1}$. The tip and frame were calibrated based on the modulus of fused silica (72 GPa) before each batch of indents in a sample. Nanoindentation consisted of batches of 500 nm and 1000 nm indents into irradiated and blanked (unirradiated) regions of the same sample at the same crystallographic orientation. This was to ensure the indentations were carried out with the same tip calibration and tip condition, sample mounting, and environmental conditions for comparing the effects of ion implantation and to avoid systematic errors. Orientation of the sample with respect to the Berkovich tip was kept constant for unirradiated and irradiated indentation. The results are averaged over 25 indents, with standard deviation error bars.

Electron backscatter diffraction (EBSD) experiments were conducted using a Zeiss Merlin FEG-SEM with a Bruker Quantax e-flash detector controlled using Bruker Esprit 2.1 software. Typical SEM conditions were 20 kV 20 nA with an acquisition time of 50 ms per pixel. Patterns were acquired with 800 x 600 pixel resolution and were all saved so that they could be analysed later using the high angular resolution EBSD code, XEBSD developed at the Department of Materials, University of Oxford, and Imperial College, London [33–35].

The procedure for analysing EBSD patterns using high angular resolution is explained fully in refs. [36,37], but will be summarised here. The simple concept is to compare EBSPs acquired from pixels in the map to a nominally unstrained reference pattern of the same crystal orientation. An applied deviatoric strain will change interplanar angles which moves Kikuchi bands in the diffraction pattern. Additionally, crystal orientation rotations will cause Kikuchi bands to move cooperatively across the screen. The diffraction pattern is segmented into 40 partially overlapping regions of interest (ROI), each of which undergoes a fast Fourier transform which is used for cross-correlation image analysis. From this a translation vector for each ROI is calculated relative to the corresponding ROI in the reference pattern. With four or more translation vectors, a self-consistent deformation tensor for the diffraction pattern can be built up with components for strain and lattice rotations [38]. Because the analysis relies on changes in orientation between lattice planes, the analysis is insensitive to the hydrostatic component of strain. The anisotropic Hooke's law can be used to determine stresses from elastic strains with elastic constants from the Materials Project database (mp-7631) [39,40]. A hydrostatic strain is then added to the strain tensor derived above to satisfy the requirement for the stress normal to the free surface to be zero[38]. It should be noted that this method is therefore unaffected by changes in stress-free lattice parameters resulting from the swelling itself, provided the swelling is isotropic.

A Zeiss Auriga FIB/SEM was used to mill a triangular cross-section micro-bridge into ion irradiated 6H-SiC, with the upper 1.2 µm being irradiated material, and the lower section being unirradiated substrate material (Figure 2). The end of this micro-bridge was cut off using sequential FIB milling and SEM imaging to release it into a cantilever. With lateral residual stress release in the irradiated layer the cantilever bent downwards, with this deflection measured using the SEM. In an unirradiated specimen, no deflection was measured. A similar technique was developed by Massl *et al.* to investigated stresses in multi-layer thin films [41,42].

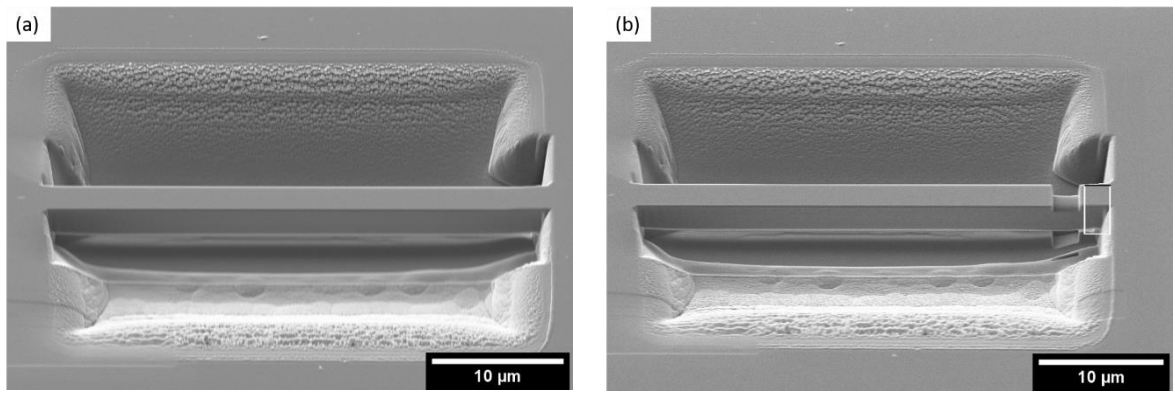

*Figure 2: (a) Micro "bridge" cut into a microcantilever, (b) right hand end being milled through, shortly prior to release.*

To quantify the compressive stress in the irradiated layer which would cause this bending, a finite element model was implemented in Abaqus. Analytical calculations based on Euler-Bernoulli beam theory do not account for relaxation of the cantilever base and lateral expansion of the cantilever is not explicitly taken into account, but may be significant based on similar experiments conducted by the group in Leoben [41,43]. To simulate the swelling of the ion implanted layer in the finite element model (FEM), a layer with the nominal thickness of the damaged layer was segmented and given a thermal expansion coefficient of 1, where the rest of the model had zero thermal expansion in the material properties. Young's modulus was 460.7 GPa, and Poisson's ratio was 0.206. These are the in-plane values derived from the anisotropic elastic constants used for HR-EBSD. A temperature was applied to simulate radiation swelling until the cantilever bent to the experimentally measured displacement with the front of the base unconstrained, the same as the experiment. The stress-free swelling strain thus derived was then used to calculate the in-plane stress and out of plane strain in the surface far from the cantilever using the same anisotropic elastic constants as were used for HR-EBSD.

Atomic force microscopy (AFM) was conducted across the blanked/irradiated boundary on the sample to investigate vertical, out of plane swelling. An Agilent 5400 AFM with a Mikro Masch NSC35 tip was operated in contact mode. The force constant of the tip was 4.5 N/m. Data was acquired using Keysight Picoview 1.20.2, and analysed using Gwyddion 2.52 [44]. Line profiles of step height are calculated as averages across the boundary of a 40 x 40 µm area map. User subjectivity during background subtraction in AFM data analysis can strongly influence the measured step heights. To minimise the influence of the user, the process here followed the guide on the Gwyddion website for measuring step heights using the "smooth bent step" function[2].

## Results

### HR-EBSD method

The area around the unirradiated-irradiated boundary was mapped by EBSD and stresses and strains were calculated by the HR-EBSD method. This allowed calculation of in-plane stress and elastic strains directly from measured diffraction patterns. The reference position for HR-EBSD calculation is taken in the lower right corner in the unirradiated material as a stress-free reference. The mean in-plane biaxial stress ($\sigma_{xx}= \sigma_{yy}$) in the irradiated region was -1.9 GPa with a mean elastic biaxial strain ($\varepsilon_{xx}= \varepsilon_{yy)}$ of -0.33%. The out of plane elastic strain $\varepsilon_{zz}$ at the surface was 0.062%, which when added to the swelling strain (= - $\varepsilon_{xx}$) gives a total out of plane strain (elastic + swelling) $\varepsilon_N^{(tot)} = 0.39\%$.

---

[2] http://gwyddion.net/survey2/step3.php

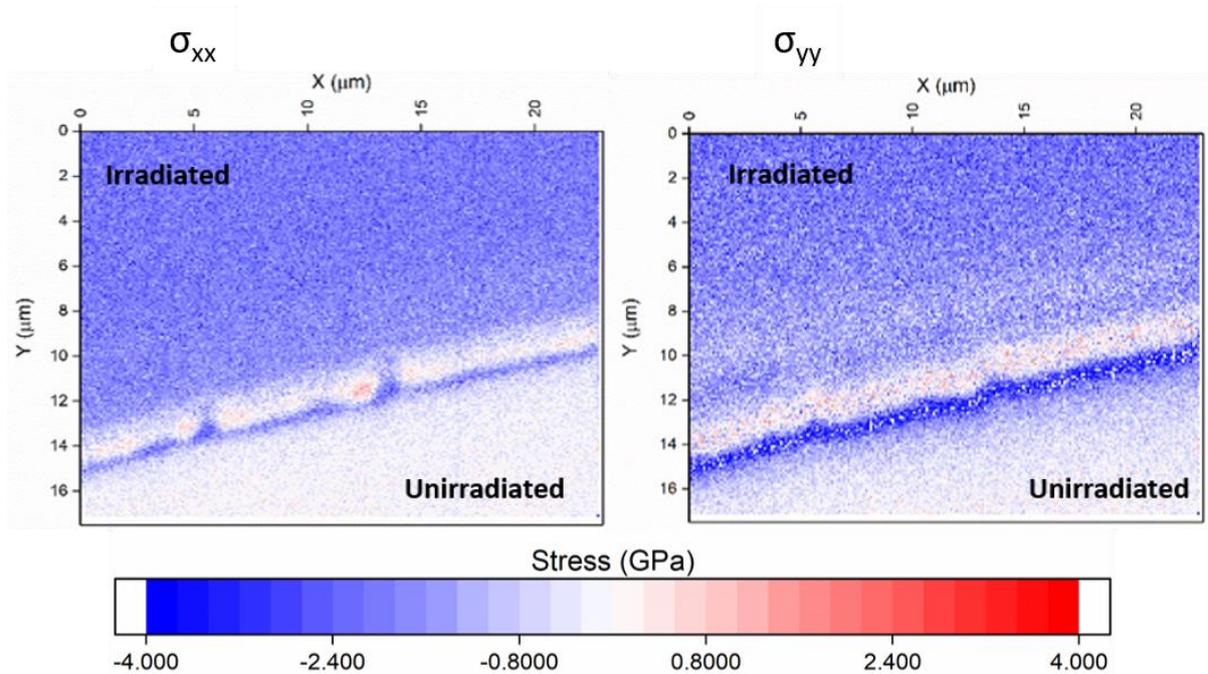

*Figure 3: In-plane HR-EBSD stress maps across the boundary between blanked and irradiated material. The stripe of light/dark contrast is an artefact of the step which caused lattice rotations.*

**Bending cantilever method**

To investigate this lateral compressive stress further, micro-bridges were made as described in the Methods section, incorporating an upper layer of radiation-damaged material, and a lower layer of the undamaged substrate. Figure 4 shows three of a sequence of images as the bridge is cut into a cantilever. (b) shows partial deflection before the final material is released in (c). The deflection of this cantilever after tilt correction is 1.92 µm, with a length of 31.8 µm, width of 3.93 µm, and height of 3.40 µm (upper 1.2 µm of damaged material, lower 2.2 µm of substrate). A video of this experiment is provided in supplementary material.

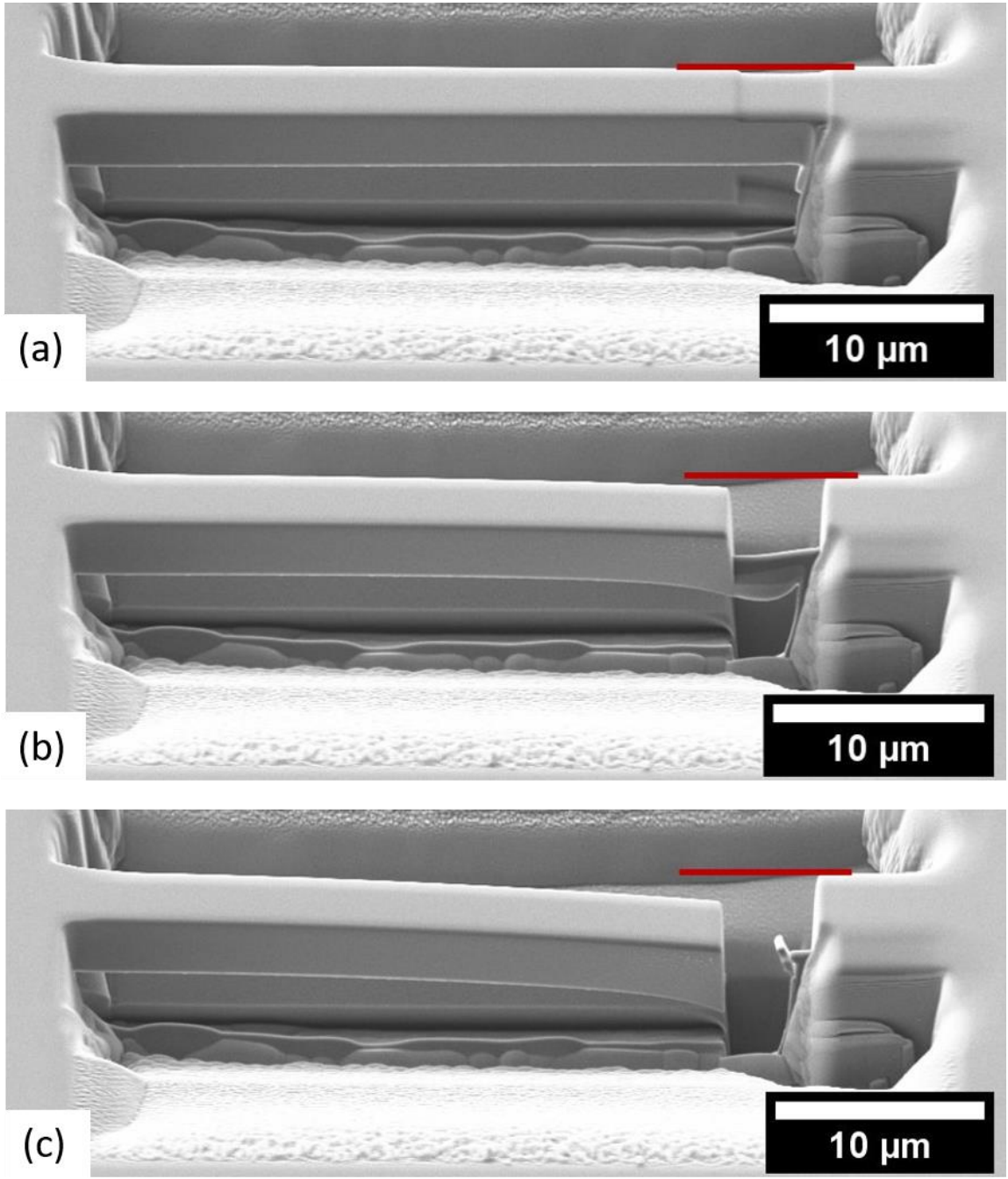

*Figure 4: Bending as the micro-bridge is cut into a microcantilever. (a) is after the first stage of milling, (b) is a partial deflection, (c) is after the cantilever has been completely freed. Red line is for reference.*

The longitudinal residual stress distribution within the FEM model of the beam is shown in Figure 5. The relaxation of the stresses in the cantilever modifies the stress distribution considerably and both the irradiated and unirradiated parts of the cross-section have tensile and compressive regions. There is a complex stress distribution around the root of the cantilever, which would be difficult to calculate analytically.

The stress-free swelling strain required to match the experimental deflection is 0.68%. This corresponds to a biaxial compressive in-plane stress far from the cantilever of -3.94 GPa.

The total out of plane expansion of the implanted layer $\varepsilon_N^{(tot)} = 0.81\%$.. The analytical solution using Euler-Bernoulli beam theory ignoring the complexities of the root gives a swelling strain of 0.72%, an error of 6%.

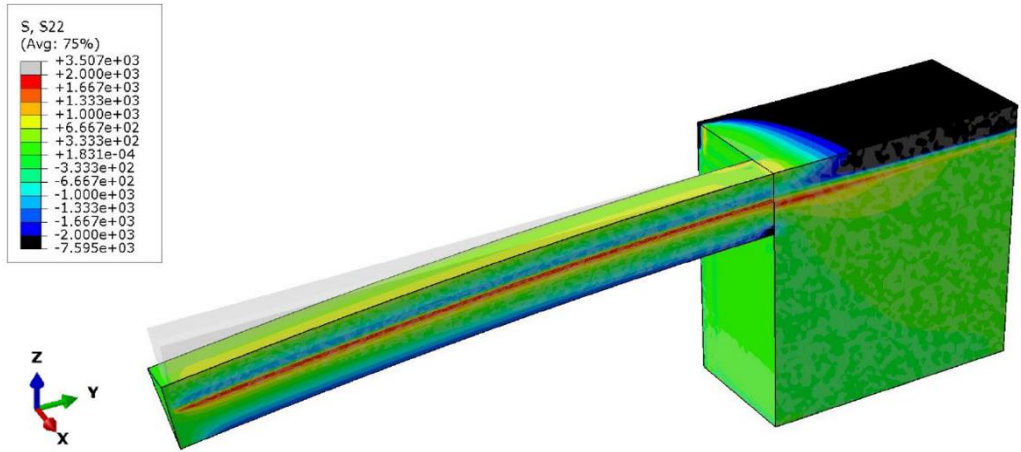

*Figure 5: FEM model of stress along the length of the beam after "swelling" Colour scale limits are set to ±2 GPa to show the stress distribution more clearly. A tensile stress is observed in the substrate at its interface with the irradiated layer. Deformation scale factor is 1× with an overlay of the undeformed beam.*

**AFM step height**

While tilting the specimen to collect EBSD patterns, a step was observed between the blanked and irradiated regions of the sample. This step was measured using AFM and found to be 18.6 nm high using the suggested analysis method for step heights (Figure 6). Over the depth of ion irradiation damage (~1.2 µm), this corresponds to an out of plane strain $\varepsilon_N^{(tot)}$ of 1.55%. Debelle & Declémy investigated lateral residual compressive stresses in ion irradiated yttria stabilised zirconia by measuring out of plane swelling using HR-XRD and derived elastic equations for stress by assuming no net in-plane strain in the irradiated layer (equation 1) [45]. As their derivation is for a cubic crystal, it is not valid in this case for hexagonal SiC. The equivalent expression for our (0001) 6H-SiC is:

$$\sigma_{//} = -\frac{C_{33}(C_{11} + C_{12}) - 2C_{13}^2}{C_{33} + 2C_{13}} \varepsilon_N^{(tot)} \qquad (1)$$

where $C_{ij}$ are the stiffness constants, $\sigma_{//}$ is the in-plane stress and $\varepsilon_N^{(tot)}$ is the total normal strain in the irradiated layer (i.e. the sum of the swelling strain, assumed to be isotropic, and the elastic strain). Using eq. (1) with the total out of plane strain measured by AFM (1.55 %) and the elastic constants from the Materials Project database [39,40] gives an in-plane biaxial stress in the implanted layer of -7.6 GPa. The corresponding stress-free swelling strain is 1.3%.

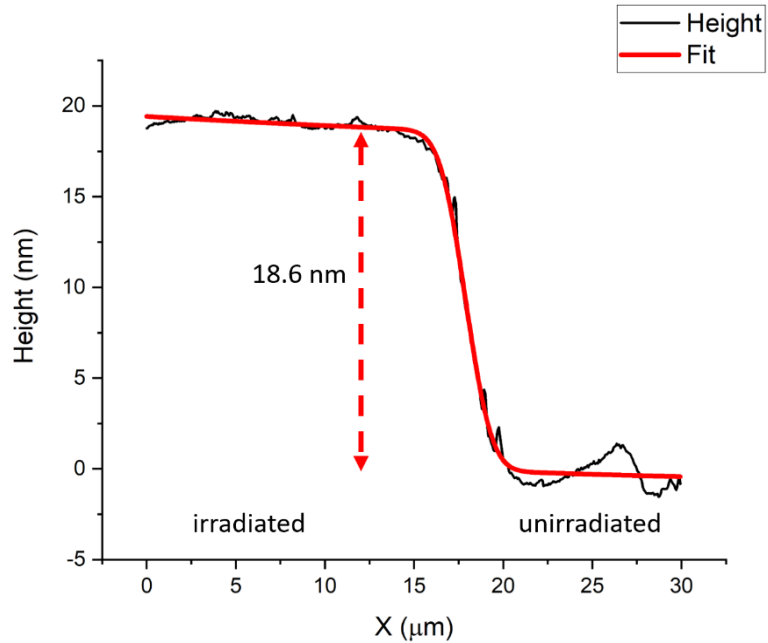

*Figure 6: AFM step height across the unirradiated-irradiated boundary. Red line is a fit to the data used for measuring the step height.*

**Nanoindentation**

Nanoindentation results show a peak of 15% increase in hardness after neon ion implantation (Figure 7). The depth of peak hardness change (550 nm) corresponds to the assumed plastic zone being entirely within the peak damaged layer according to the SRIM damage profile. Below 50 nm nanoindentation data is unreliable due to tip contact effects and is omitted for clarity. Near the surface (<100 nm) nanoindentation is dominated by the indentation size effect and the hardness change is small until the sampling volume is sufficiently large to be representative of the specimen. Significantly, there is no surface radial fracture from indents in the irradiated material. This affects the indentation deformation mechanism and the extracted mechanical properties and will be explored further in a future paper.

Elastic modulus has a maximum reduction of -12% which returns towards the unirradiated value with increasing indentation depth. Elastic modulus is a long-range effect and the undamaged substrate is having an effect on this measurement as the irradiated modulus returns towards the unirradiated modulus.

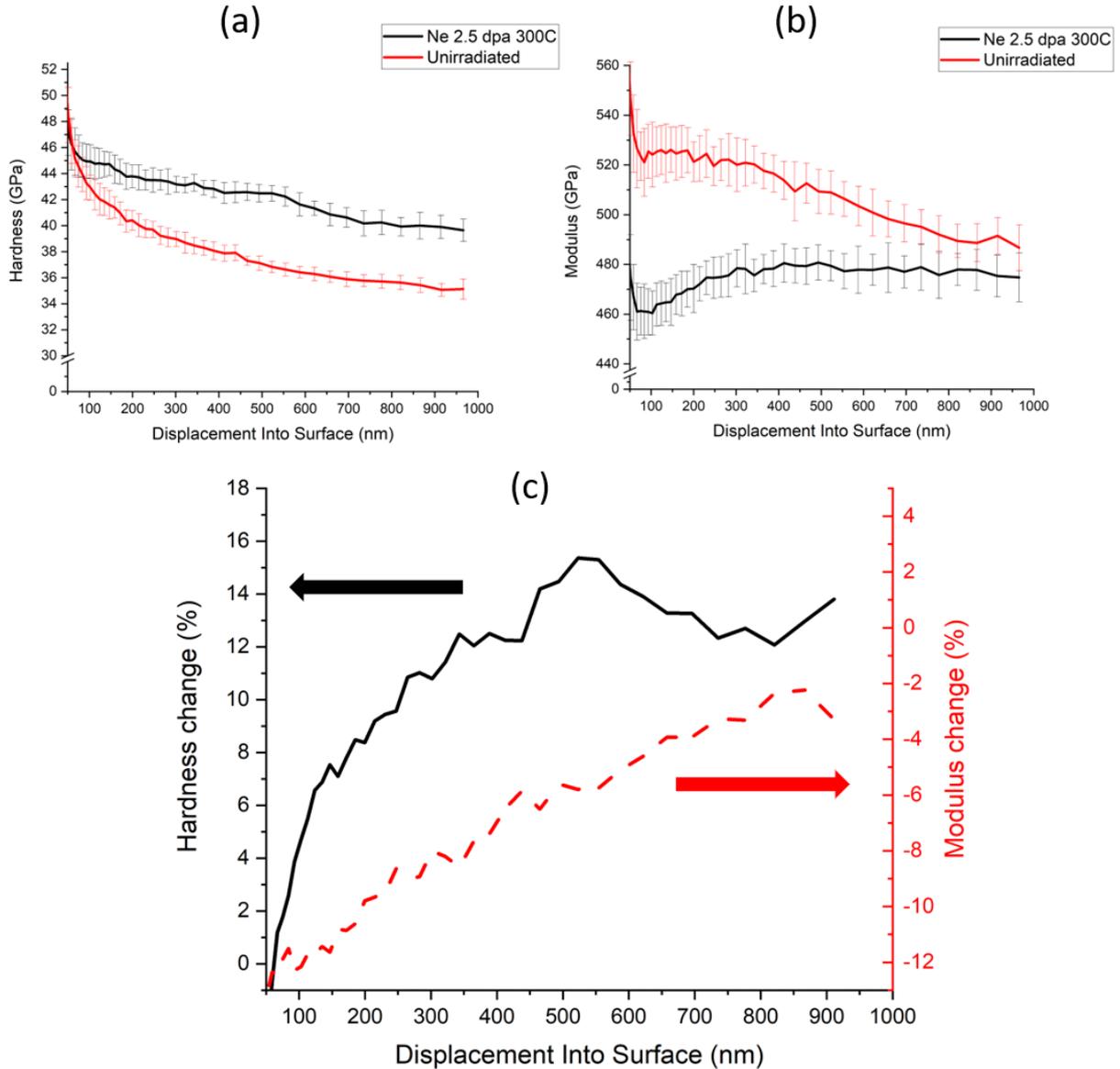

*Figure 7: Hardness (a) and modulus (b) as a function of depth for irradiated and unirradiated indents. (c) Change in nanoindentation hardness and modulus of 6H-SiC after 2.5 peak dpa neon ion irradiation*

**Discussion**

**Ion irradiation-induced residual stress**

It is well known that radiation damage causes swelling in silicon carbide [46]. The origin of this radiation-induced strain is lattice swelling due to defects in the sample. Li *et al.* modelled the excess volume of point defects in SiC, finding that all except the $C_{Si}$ anti-site defect have a positive volume [47]. Additionally, swelling can arise from bubbles and voids at temperatures over 1000°C [48]. At the irradiation temperature of this work, defect swelling is expected to be in the point defect swelling regime. As neutron irradiation (approximately) uniformly damages the bulk of a sample, defects are free to swell the specimen in all dimensions unless there are engineering constraints or significant crystallographic anisotropy. Bulk dimensional measurements and XRD of neutron irradiated polycrystalline 3C-SiC shows uniform dimensional and lattice expansion in all directions [46]. At low temperatures single crystal 6H-SiC swells identically to CVD 3C-SiC under neutron irradiation [49].

In ion irradiation experiments, a thin layer is irradiated on a thick substrate. In this case, the damaged layer is ~1.2 µm thick on a 300 µm substrate. This stiff substrate constrains lateral swelling, resulting in a lateral compressive residual stress. The free surface is allowed to expand in $\varepsilon_{zz}$, as is observed by AFM step measurements and diffraction experiments both here and in the literature [26,50–55].

The in-plane stress values calculated by the HR-EBSD, bending cantilevers, and AFM step height are of the same order of magnitude, but do not quantitatively agree. EBSD patterns arise from near the specimen surface (~40 nm depth). This corresponds to a region of lower irradiation damage, nominally 0.9 dpa from the SRIM calculation (Figure 1), but residual damage may be lower than this due to the free surface acting as a defect sink. Although peak damage is 2.5 dpa, the average damage in the layer is 1.8 dpa.

Linear scaling of strain with dose has previously been observed by XRD strain analysis of helium implanted SiC in ref [55]. Scaling the stress and stress-free swelling strain calculated by HR-EBSD (-1.9 GPa and 0.33% respectively at 0.9 dpa) to the mean damage in the irradiated layer (1.8 dpa) gives a stress of -3.8 GPa and a swelling strain of 0.66%, closely matching the corresponding values of -3.9 GPa and 0.68% estimated using the bending cantilever method.

The stress calculated from the AFM step method is significantly higher than the estimates from the other two methods. One reason for this is that measuring step heights using AFM is strongly dependent on background subtraction during the data analysis. Although an agnostic method of background subtraction was used in this work, it may not be correct. Background subtraction based on user judgment gave a step height of 14 nm, corresponding to $\varepsilon_N^{(tot)} = 1.2\%$ and corresponding biaxial stress from equation (1) of -5.7 GPa. This method is also particularly sensitive to some of the assumptions of the analysis. One of these, discussed further in the paragraph below, is the assumption that the swelling is isotropic. The HR-EBSD method also relies on this assumption but in measuring strains in a range of directions is arguably less sensitive to it. The cantilever method is unaffected by this consideration because it is an absolute measurement of in-plane strain and stress. Another assumption of the AFM method is that the elastic properties of the irradiated layer are the same as those in the undamaged material. The nanoindentation results show that the irradiation damage actually reduced the stiffness by up to 12%. If accounted for, this would reduce the stress from the AFM method to around 5 GPa. Although this would also affect the HR-EBSD method in principle, the shallow depth sampled here was less damaged, as noted above. This would also affect the detailed conclusions from the cantilever method, but again less so because the stress measured is partly deduced from the response of the undamaged layer. In summary, the AFM step measurement technique appears to be the least reliable method due to the potential for user influence during analysis and its susceptibility to the assumptions it involves.

A consequence of this constraint of the implanted layer and the resulting compressive stress which needs to be investigated further is the effect it has on fundamental radiation defects. Does it favour the formation of "small" defects? Does it hinder the growth of larger structures? Are some defect structures created to try to accommodate the volume of defects and minimise energy? Such effects may also lead to anisotropic swelling. Observations of chemical defects in ion irradiated layers by Raman spectroscopy is not consistent with those observed in neutron irradiated SiC [27,28,56]. In particular the small shift in the Si-C peak position after ion implantation suggests constrained bond length expansion, leading to lateral residual compressive stresses. Compressive stresses may favour defects with smaller excess volumes, possibly $C_{Si}$ anti-sites or $V_{Si}$ silicon vacancies [47]. Recent work with high resolution XRD on single energy ion implanted SiC has shown a strain gradient which enhances diffusion of interstitial atoms created near the surface towards more highly strained regions of material at the peak damage depth. This leads to different defect structures such as helium and void platelets [53,54]. This strain gradient effect was also seen in annealing experiments where defects migrated towards the highly strained region at the damage peak at a lower energy than predicted [52]. Observing the influence of this stress on defects using traditional TEM techniques is impossible as producing a thin TEM foil necessarily relieves residual stresses; analysis of stress relaxation in TEM lamella preparation using convergent-beam electron diffraction can be used to evaluate the residual stress in the original material [57].

**Effect of residual stress on material properties**

Large hardening has been observed for ion irradiated 6H-SiC in the literature, however this is not consistent with the observation of no hardening in neutron irradiated 6H-SiC in Chen *et al*'s work [21]. Additionally, Chen *et al.* observed a larger reduction in modulus in neutron irradiated 6H-SiC than is observed after ion irradiations [21]. Their irradiation conditions were different to this work, but it is the only literature with a systematic comparison of neutron and ion irradiation effects on nanoindentation in 6H-SiC. They attribute the hardening to lateral compressive stresses caused by radiation swelling. Ion implantation causes a smaller reduction in modulus than neutron irradiation [21]. Elastic modulus should not be affected directly by residual stress; it is related to covalent bond density and stiffness (interatomic potential energy) which is reduced when radiation defects break bonds and damage the crystal lattice. Implanted ions can substitute into the lattice to re-form bonds to some extent, possibly explaining the smaller reduction in elastic modulus than in neutron irradiated materials. The ability to chemically substitute into a SiC lattice will likely depend on the implanted ion species and is explored more in [27]. Elastic measurements by nanoindentation are inherently long-ranged, and include a contribution from the undamaged substrate, as shown by the return towards unirradiated values of elastic modulus with deeper indentation. The contribution of unirradiated material to elastic modulus measurements at shallow depths is unclear, but it should not be observed in neutron irradiations.

Based on their nanoindentation measurements, Chen *et al.* investigated the lateral residual stress using FEM simulations of nanoindentation in irradiated 6H-SiC, adjusting a compressive stress in a damaged layer until simulated load-displacement curves matched the experimental load-displacement curves [21]. They found compressive biaxial stresses between 5-13 GPa, which are similar in magnitude to those measured in this work. Their ion irradiations were at low temperature (~50°C) which causes more swelling and more associated residual stress than in this work. This approach of matching experimental and FEM load-displacement curves ignores the effect of suppressed fracture and any effect of radiation defects in the experimental nanoindentation measurements.

Even in ductile metals such as iron-chromium alloys, anomalous micromechanical results after ion irradiation have tentatively been attributed to lateral residual stresses in ion implanted layers [58]. The effect of residual stress is perhaps not as obvious in metals as in SiC as indents in neither the unirradiated, nor irradiated metal crack; there is not an obvious change in deformation response which would spark further investigation. Ion implantation into a thin sample of austenitic stainless steel resulted in curvature which was measured by optical profilometry [59]. Based on this curvature a residual stress of -250 MPa was calculated. Subsequently these authors developed a technique using nanoindentation coupled with FEM to measure residual stress in this material, and found a compressive stress of -631 MPa in an ion implanted sample with a thick substrate which did not bend [60].

In the context of nuclear materials, ion implantation is intended to introduce displacement damage to simulate the effects of neutron irradiation displacement damage. Nanoindentation is intended to measure any changes caused by this displacement damage.

However, for other applications ion implantation is used as an intentional surface treatment to strengthen and improve the wear resistance of materials [61,62], and nanoindentation can be used to measure biaxial residual stresses [63]. When using nanoindentation to study changes to mechanical properties caused by radiation defects, it is important to also consider residual stresses introduced by ion irradiation.

**Conclusions**

Compressive biaxial residual stress has directly been observed in ion implanted 6H-SiC, and is contributing to the large hardening measured by nanoindentation. The stresses measured in this work were several GPa in magnitude. The origin of the compressive biaxial stress is radiation-induced swelling constrained by the undamaged substrate. This constrained swelling effect appears to have a significant (non-negligible) effect on micromechanical properties and may also influence the radiation defects formed. This effect should be taken into account when measuring micromechanical properties of all ion irradiated materials which are susceptible to radiation-induced swelling, including metals. As this effect is due to the undamaged substrate and surrounding material constraining swelling in the plane of the specimen, any lift-out type analysis technique such as TEM is likely to be seeing a different structure to what is caused by the ion implantation, and which is present during measurements such as XRD, SEM, spectroscopy, or nanoindentation. In this case it is important to be able to investigate structural damage in a bulk specimen without the need for nanoscale TEM specimens.

The bending of microcantilevers FIBed into the implanted surface is a reliable method of measuring the residual stresses and is less affected by the assumptions involved than most methods. HR-EBSD gave stresses that can be reconciled with those from microcantilever measurements but can only sample a very thin surface layer, where irradiation damage is lower than deeper into the sample. The measurement of surface step heights was less reliable owing to experimental uncertainties and susceptibility to assumptions involved in the analysis.

In an operating fusion or fission reactor, components will be irradiated at different temperatures and to different doses, resulting in different swelling. This will cause significant residual stresses in components during operation of the reactor, which must be considered and accounted for in the design, along with degraded mechanical and thermal properties.

**Acknowledgements**

Thanks to Dr Nianhua Peng at Surrey Ion Beam Centre for running the ion implantation, Nicola Flanagan at Oxford Materials Characterisation Service for conducting the AFM measurement, and Prof. Edmund Tarleton for assistance with the FEM model. Financial support through the EPSRC Science and Technology of Fusion CDT [grant number EP/L01663X/1] is gratefully acknowledged. The authors acknowledge use of characterisation facilities within the David Cockayne Centre for Electron Microscopy, Department of Materials, University of Oxford, alongside financial support provided by the Henry Royce Institute (Grant ref EP/R010145/1). This work has been carried out within the framework of the EUROfusion Consortium and has received funding from the Euratom research and training programme 2014-2018 and 2019-2020 under grant agreement No 633053. The views and opinions expressed herein do not necessarily reflect those of the European Commission.